\documentclass[twocolumn,showpacs,preprintnumbers,amsmath,amssymb]{revtex4}
\usepackage{graphicx}% Include figure files
\usepackage{dcolumn}% Align table columns on decimal point
\usepackage{bm}% bold math

\begin{document}

\title{Charge transfer via a two-strand superexchange bridge in DNA}
\author{X.F. Wang and Tapash Chakraborty}
\affiliation{Department of Physics and Astronomy, The University of
Manitoba, Winnipeg, Canada, R3T 2N2}

\begin{abstract}
Charge transfer in a DNA duplex chain is studied by constructing a
system with virtual electrodes connected at the ends of each DNA
strand. The system is described by the tight-binding model and its
transport is analyzed by the transfer matrix method. The very weak
distance dependence in a long (G:C)(T:A)$_{\cal M}$(G:C)$_3$ DNA chain
observed in experiment [B. Giese, {\it et al.}, Nature {\bf 412},
318 (2001)] is explained by a unistep two-strand superexchange
bridge without the need for the multi-step thermally-induced hopping
mechanism or the dephasing effect. The crossover number ${\cal M}_c$
of the (T:A) base pairs, where crossover between the strong and weak distance
dependence occurs, reflects the ratio of intra- and inter-strand
neighboring base-base couplings.
\end{abstract}
\maketitle
In recent years, charge migration in DNA has attracted considerable interest 
among the physics, chemistry, and biology communities. Charge transfer in 
DNA is important for functioning of molecular electronic devices 
\cite{erez} as well as in understanding the DNA oxidative damage and 
repair \cite{oxidative}. Additionally, DNA offers a platform for 
fundamental physical understanding of systems in the nano-scale. It has 
been a long-standing problem to understand whether the charge transfer in 
DNA occurs via a unistep coherent superexchange process or a multi-step 
incoherent thermally-induced hopping process \cite{book2}. In a one-strand 
unistep model the transfer rate is exponential and is strongly distance
dependent \cite{mcco, muji, berl}. The multi-step hopping model on the other 
hand, predicts a weak dependence on the distance. Both of these ideas have 
received experimental supports \cite{kell,book2}. Recent experiments have 
shown that the sequence of base-pairs may account for the transition between 
the strong and weak distance dependence of charge rates in DNA, but the 
underlying mechanism is not yet clear \cite{lewi, gies}. In 
Ref.~\cite{gies}, the transfer rate through a DNA of sequence 
(G:C)(T:A)$_{\cal M}$(G:C)$_3$ was measured for different ${\cal M}$. 
The charge transfer showed a strong distance dependence when ${\cal M}<3$, 
but almost no distance dependence for ${\cal M}>3$. To explain this 
distance-dependence crossover at ${\cal M}=3$, a combination of
coherent superexchange and a hopping mechanism (incoherent) -- the 
variable-range hopping model, was proposed to allow for a transition 
between these two regimes \cite{berl,reng}. In the former process, 
the donor and the acceptor of the charge are coupled to the bridge of 
higher energy, without any chance of intermediate relaxation. The charge 
remains in a quantum state over the bridge that works as a tunneling 
barrier. In the hopping process, relaxation is introduced into each site 
and the charge loses its coherence (phase) when it reaches a site.  
A population parameter for each site was necessary
to describe the distribution of the charge over the bridges. 

In this letter, we demonstrate that the experimentally observed 
distance-dependence crossover can as well be explained by a simple 
two-strand superexchange model. The almost zero distance dependence of the 
charge transfer at a long chain is shown to be a result of the inter-strand 
coupling in the DNA. In this model, the system still remains coherent 
and the charge transfer occurs in a unistep way.

\begin{figure}
\begin{center}
\begin{picture}(330,130)
\put(0,5){\includegraphics{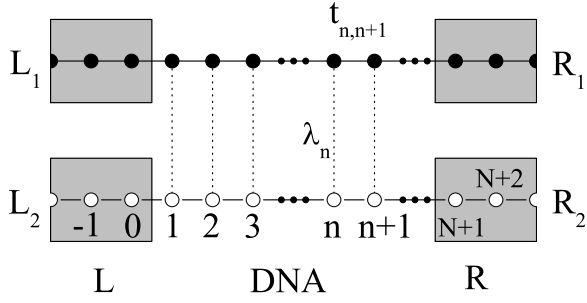}}
\end{picture}
\vspace{-1.0cm}
\protect\caption{Schematic illustration of the system. The first strand
(filled circle) has a DNA base sequence G(T)$_{\cal M}$GGG and the second
strand (empty circle) a sequence C(A)$_{\cal M}$CCC. The four gray areas
indicate the four virtual electrodes connected to the DNA chain. Current
is injected into the first strand through the left electrode L$_1$ and
measured at the right electrode R$_1$.}
\label{fig:fig1}
\end{center}
\end{figure}

We consider a DNA duplex chain of $N$ Watson-Crick base pairs
connected to four semi-infinite one-dimensional (1D) electrodes with one 
for each end of the first and the second strand as illustrated in 
Fig.~\ref{fig:fig1}. The tight-binding Hamiltonian of the system is
%\begin{widetext}
\begin{eqnarray*}
H &=&2\sum_{n=-\infty}^\infty
[\varepsilon _n c_n^\dag c_n
-t_{n,n+1} (c_n^\dag c_{n+1}+c_{n+1}^\dag c_n)]\\
&+&2\sum_{n=-\infty}^\infty [u _n d_n^\dag d_n
-h_{n,n+1}(d_n^\dag d_{n+1}+d_{n+1}^\dag d_n)] \\
&-&2\sum_{n=1}^N
\lambda_n (c_n^\dag d_n+d_n^\dag c_n).
\label{eq:hamiltonian}
\end{eqnarray*}
%\end{widetext}
Here $c_n^\dag$ ($d_n^\dag$) is the creation operator of holes
in the first (second) strand on site $n$ of the DNA chain (for
$1\leq n \leq N$), the left electrodes ($n\leq 0$), and the right
electrodes ($n\geq N+1$). The on-site energy of site $n$ in the
first (second) strand is denoted by $\varepsilon_n$ ($u_n$), which
is equal to the highest occupied molecular orbit (HOMO) energy of
the base on this site in the DNA chain and the center of conduction
band in the electrodes. The coupling parameter of the first (second)
strand $t_{n,n+1}$ ($h_{n,n+1}$) is equal to the intra-strand
coupling parameter $t_d$ between neighboring sites $n$ and $n+1$
of the DNA for $1\leq n \leq N-1$, one-fourth of the conduction band-width
in the electrodes $t_m$ for $n \leq -1$ and $n\geq N+1$, and the
coupling strength $t_{dm}$ between the electrodes and the DNA strands
for $n=0$ and $n=N$. The inter-strand coupling between sites in the
same Watson-Crick base pair is described by $\lambda_n$. The factor
2 multiplied to each sum in Eq.~(\ref{eq:hamiltonian}) arises from
the spin degeneracy.

%Fig. 1

In what follows, we have studied the intra-molecular hole transfer property
along the DNA duplex chain after charges are injected (optically or
electrically) into the base on site 1 of the first strand. To facilitate
our calculation, we connect one virtual electrode to the left end of each
DNA strand as the injector and another to the right end as the drain for
holes. To minimize the contact effect, we assume a strong coupling (of
coupling parameter $t_{01}=t_{N,N+1}=h_{01}=h_{N,N+1}=t_{dm}=1.5$ eV) between
the electrodes and the sites at the ends of the DNA strands,
and choose a band width ($4t_m$) in the electrodes such that the optimal
injection condition $t_d\times t_m=t_{dm}^2$ \cite{maci} is satisfied.
Our result is independent of the choice of the value of $t_{dm}$
once it is much larger than the coupling parameter between the sites
inside the DNA. In this case, the added electrodes does not become
a bottleneck of the system for the charge transfer and the calculated
result predominantly reflects the properties of the DNA chain.

The transport properties are evaluated by the transfer matrix method
\cite{maci,carp}. For an open system, the secular equation 
is expressed as a group of infinite number of equations of the form
\begin{eqnarray*}
t_{n-1,n}\Psi_{n-1}+(\varepsilon_n-E)\Psi_n
+\lambda_n \Phi_n+t_{n,n+1}\Psi_{n+1}=0\\
h_{n-1,n}\Phi_{n-1}+(u_n-E)\Phi_n
+\lambda_n \Psi_n+h_{n,n+1}\Phi_{n+1}=0
\end{eqnarray*}
with $\Psi_n$ ($\Phi_n$) the wave function of the first (second)
strand on site $n$. The wave functions of the sites $n+1$ and $n$ are
related to those of the sites $n$ and $n-1$ by a transfer matrix $\hat{M}$,
\begin{equation}
\left( \begin{array}{c}
\Psi_{n+1}\\
\Phi_{n+1}\\
\Psi_n\\
\Phi_n
\end{array}\right) =\hat{M}
\left(
\begin{array}{c}
\Psi_n\\
\Phi_n\\
\Psi_{n-1}\\
\Phi_{n-1}
\end{array}\right),\,\,\,
\label{eq:matrix}
\end{equation}
with
%\begin{widetext}
\begin{equation*}
\hat{M}=
\left[ \begin{array}{cccc}
\frac{(E-\varepsilon_n)}{t_{n,n+1}}& \frac{-\lambda_n}{t_{n,n+1}}&
-\frac{t_{n-1,n}}{t_{n,n+1}}&0\\
\frac{-\lambda_n}{h_{n,n+1}}&\frac{(E-\varepsilon_n)}{h_{n,n+1}}
&0&-\frac{h_{n-1,n}}{h_{n,n+1}}\\
1&0&0&0\\
0&1&0&0
\end{array}\right].
\end{equation*}
%\end{widetext}
The transmission can be calculated by assuming the plane waves
propagating in the electrodes. Here we are interested in
the case where only holes are injected from electrode L$_1$
to the first strand. The hole wave functions in the L$_1$ electrode
is $\Psi_n=(A e^{ik_Lna}+B e^{-ik_Lna})$ ($n\leq 0$) and in the
R$_1$ electrode $\Psi_n=C e^{ik_Rna}$ ($n\geq N+1$). The distance 
between two neighboring bases along any DNA strand is $a=3.4$ \AA. 
Using Eq.~(\ref{eq:matrix}), we express the output wave amplitude $C$ 
in terms of the input wave amplitude $A$ and evaluate the transmission 
to R$_1$ electrode as
$$T(E)=\frac{|C|^2\sin(k_R a)}{|A|^2\sin(k_L a)}.$$
We have chosen the normalized incident amplitude to be
$A=1/\sqrt{|\sin(k_L a)|}$.

To evaluate the transfer rate or current of a charge (hole)
from the donor at the left-end site to the acceptor at the right-end
site of the first strand, we need to know the chemical potential at
each end. In the experiment of Ref.~\cite{gies}, a hole was
injected to the left-end site. This means that the left chemical
potential is approximately the on-site energy of this site while the
right one is less. During the charge transfer process, the hole may
retain the same energy if no inelastic scattering occurs or lose energy
via the electron-phonon scattering or other inelastic collisions.
Here we do not deal with these inelastic scattering mechanisms explicitly
but analyze two limiting situations, between which the real charge
transfer process occurs. Since our results for the distance dependence
of the transfer rate from the two limits converge (see below), we
conclude that our results are reliable.

In the first limit, we assume that there is no inelastic scattering
involved and the hole energy is conserved during the transfer process.
The transfer rate is proportional to the conductance of the system at
equilibrium. For a small electric potential difference $k_B T/e$,
the current is
\begin{equation}
I=\frac{2e^2}{h}\int_{-\infty}^\infty
dE\, T(E)[1-f(E)]f(E).
\label{eq:i1}
\end{equation}
Here
$f(E)=1/\exp[(E-\mu)/k_B T]$ is the Fermi function. The room temperature 
$T=300$ K is assumed and the on-site energy of site 1 in the first strand 
is used as the chemical potential $\mu$. In the second limit, we assume 
that the hole can lose energy freely during the process, and the transfer 
rate is proportional to the total current via all channels of energies 
below the hole's initial energy. This corresponds to an infinitely low 
chemical potential at the right electrode and the current is
\begin{equation}
I=\frac{2e^2}{h}\int_{-\infty}^\infty
dE\, T(E)f(E).
\label{eq:i2}
\end{equation}

We now calculate the distance dependence of the transfer rate using 
Eq.~(\ref{eq:i1}) and Eq.~(\ref{eq:i2}) in a DNA duplex, where the first 
strand has the base sequence G(T)$_{\cal M}$GGG as in the experiment of 
Ref.~\cite{gies}. The HOMO energies for bases G, C, T, A, are $E_G=7.75$, 
$E_C=8.87$, $E_T=9.14$, and $E_A=8.24$ eV respectively \cite{maci}. A 
uniform intra-strand hopping parameter $t_{n,n+1}=h_{n,n+1}=t_d$ ($1\leq n 
\leq N-1$) and  a uniform inter-strand hopping parameter $\lambda_n=\lambda_d$ 
($1\leq n \leq N$) between any two neighboring bases in the DNA are used.

\begin{figure}
\begin{center}
\begin{picture}(450,220)
\put(-15,25){\includegraphics{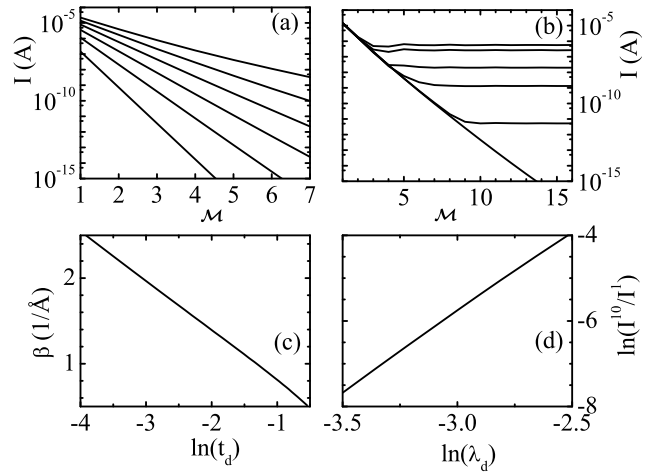}}
\end{picture}
\vspace{-1.4cm}
\protect\caption{(a) Current $I$ versus ${\cal M}$ for $t_d=0.1, 
0.2, 0.3, 0.4, 0.5, 0.6$ eV (from lower to upper curves) for zero
inter-strand coupling. The displayed results are from Eq.~(\ref{eq:i2}) 
and identical results are obtained from Eq.~(\ref{eq:i1}) in all the 
panels. (b) Same as in (a) at fixed $t_d=0.5$ eV but for
$\lambda_d=0, 5, 20, 40, 80, 100$ meV corresponding to curves counted
from the bottom. (c) The $\beta$ value calculated from the slope of
the lines in (a) versus $\ln t_d$. (d) $\ln (I^{10}/I^1)$, where 
$I^{\cal M}$ is the current for a chain with ${\cal M}$ (A:T) base 
pairs, versus $\ln \lambda_d$. The unit of $t_d$ and $\lambda_d$ is eV.}
\label{fig:fig2}
\end{center}
\end{figure}

First we switch off the inter-strand coupling and calculate the
dependence of the current $I$ on ${\cal M}$ as shown in Fig.~\ref{fig:fig2}
(a), for different values of the intra-strand coupling parameter $t_d$.
We find an exponential dependence of the current
\begin{equation}
I=I^{\cal M} \propto e^{-\beta {\cal M}a}.
\label{eq:current}
\end{equation}
We then extract the values of $\beta$ for different $t_d$ and plot in
Fig.~\ref{fig:fig2}(c) as $\beta$ vesus $\ln(t_d)$ calculated via Eq.~(\ref{eq:i2}).
The curves are almost linear, very similar to the results of Eq.~(\ref{eq:i1}),
and converge to the approximate formula \cite{mcco,muji,berl}
\begin{equation}
\beta=\frac2a\ln\frac{t_d}{E_T-E_G}.
\label{eq:beta}
\end{equation}
This is the well-known 1D superexchange result in the literature and
has been derived in many different ways. This agreement
confirms the validity of our model.

In the next step, we fix $t_d$ and switch on the inter-strand coupling
by varying $\lambda_d$. The result is displayed in Fig.~\ref{fig:fig2}(b)
where we choose $t_d=0.5$ eV and plot $I$ versus ${\cal M}$ for a
series of $\lambda_d$. Note that the charge transfer occurs via 
$\pi$-electrons and generally $\lambda_d < t_d$ \cite{voit}. For finite 
$\lambda_d$, the current drops exponentially with increasing ${\cal M}$ 
for small ${\cal M}$ and then becomes almost flat with oscillations 
around a limiting current $I^\infty$ for large ${\cal M}$. The crossover number
${\cal M}_c$ depends on the strength of the inter-strand coupling parameter. 
The weaker the inter-strand coupling is, the bigger the ${\cal M}_c$. 
The dependence of $I^\infty$ on $\lambda_d$ is approximately illustrated in
Fig.~\ref{fig:fig2}(d), where the normalized current $I^{10}/I^1$ of the DNA
chain at ${\cal M}=10$  is plotted versus $\ln(\lambda_d)$. Again, two almost
identical straight lines are found corresponding to the two limiting situations
based on Eq.~(\ref{eq:i1}) and Eq.~(\ref{eq:i2}) and can be approximately
expressed as
\begin{equation}
\ln (I^{10}/I^1)=5.7+3.9\ln (\lambda_d).
\label{eq:limit}
\end{equation}
From Eqs.~(\ref{eq:current})-(\ref{eq:limit}), we estimate the ratio
of inter- and intra-strand coupling from the crossover number 
${\cal M}_c$. Since the environment can change $\lambda_d/t_d$, we 
predict that the transition number may vary and be different from 3 
when the experimental environment changes.

%Fig. 2

Calculating the current $I$ before and after adding a (T:A) base 
pair at site $n$ with zero or nonzero inter-strand coupling 
$\lambda_n$, we find that the distance-dependence crossover has a
topological origin, viz., from the 1D chain charge transport to a partly 
two-dimensional (2D) network. When a new (T:A) base pair is inserted 
into the DNA chain, a new superexchange channel is opened through its 
inter-strand coupling and the corresponding contribution exactly compensates 
the loss of charge transfer rate that would incur because of an extra barrier to 
the existing channels.

\begin{figure}
\begin{center}
\begin{picture}(450,200)
\put(0,20){\includegraphics{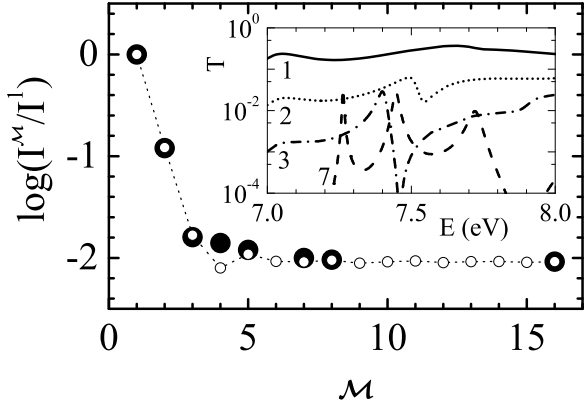}}
\end{picture}
\vspace{-1.4cm}
\protect\caption{Normalized transfer rate measured in Ref.~\cite{gies} 
(filled circle) and theoretical fit using this model (open circle), 
$\log (I^{\cal M}/I^1)$, are plotted as functions of ${\cal M}$ (T:A) 
base pairs between the (G:C) and the triple (G:C)
base pairs. Inset: The corresponding transmission $T$ versus
energy $E$ for ${\cal M}=1$ (solid line), 2
(dotted line), 3 (dot-dashed line), and 7 (dashed line).}
\label{fig:fig3}
\end{center}
\end{figure}

In Fig.~\ref{fig:fig3}, we fit the ${\cal M}$ dependence of the charge
transfer rate observed in Ref.~\cite{gies} using intra- and inter-strand
coupling parameters $t_d=0.52$ eV and $\lambda_d=0.07$ eV respectively.
Eq.~(\ref{eq:i2}) is employed in the calculation. The agreement between
the experimental and theoretical results are very good except that a small
oscillation is visible in the theoretical result near ${\cal M}_c$. This
oscillation results in the deviation of the empty circle from the filled
circle at ${\cal M}=4$. When Eq.~(\ref{eq:i1}) is used, similar result
is obtained but with a stronger oscillation. The oscillations reflect
the fact that we have treated the system as a coherent system by neglecting
the dephasing effect from the environment and the relaxation process
from phonons.

To get a clear picture of the process, we plot as inset in Fig.~3, the 
transmission $T$ as a function of the hole energy $E$ for systems with 
${\cal M}=1$, 2, 3, and 7 in an energy range near and below the $G$ base 
HOMO energy $E_G$. In the $T$ spectrum, each peak represents a transport 
channel and there are more fine structures or peaks when more base pairs 
are added to the system. When ${\cal M}$ varies from 1 to 3, the 1D chain 
transport dominates and only one principal transmission peak is important. 
The principal peak shifts when ${\cal M}$ varies due to the shift of 
energy of the channel; its height drops rapidly leading to an 
exponential decrease of charge transfer rate.
%the transmission drops rapidly overall because the contribution
%from inter-strand coupling is not comparable to that from intra-strand coupling. 
%For small ${\cal M}$, the barrier between the (G:C)
%base pair and the triple (G:C) base pairs is approximately the %one-strand
%bridge because the modification of inter-strand coupling is %negligible.
If we add more (T:A) base pairs to the DNA duplex, the principal $T$ peak 
drops to a level comparable to that of other peaks and results in a 
corssover from 1D chain transport to 2D network transport.
% show the the barrier profile
%is in such a form that holes began to tunnel through the two-strand bridge
%rather than the one-strand G(T)$_{\cal M}$GGG bridge. The overall transmission
%does not drop but shows more fine structures or oscillations. Each transmission
%peak along the energy axis is a transmission window of charge transfer and
%these windows shift to different energies as the base-pair sequence changes.
In the absence of any inelastic scattering
% and when the energy of the hole
%is conserved during the charge transfer process, 
the charge transfer rate versus ${\cal M}$ oscillates as a result of the 
energy shift of the transport channels and the energy conservation of the charge.
%corresponding to the hole energy. 
With the assistance of the phonon, however,
% in the presence of inelastic
%coupling such as the electron-phonon interaction, a hole 
the charge can use channels of energy different from its initial energy and 
phonons may play an important role in assisting the charge transfer.

%Fig. 3

Our intra-strand coupling parameter $t_d=0.52$ eV used to fit the
measurement is consistant with the GG coupling parameter extracted
from a direct I-V measurement through a DNA of 30 (G:C) base pairs
\cite{pora}. This fit parameter is much larger than the
{\it ab initio} values \cite{maci,voit}. The reason of this disagreement 
is not yet clear but it may be related to the electron-phonon interaction.
This interaction may also affect the temperature dependence of the 
charge transfer through the system \cite{barb}. In this work, 
we do not treat the electron-phonon interaction in detail but we 
expect that these interaction can determine the position of the real 
current between our two current limits [Eq.~(\ref{eq:i1}) and Eq.~(\ref{eq:i2})] 
discussed above. The dephasing effect due to the environment can also 
result in a weak distance dependence for a 1D long bridge system \cite{muji},
which is a different mechanism from what we propose here. The dephasing
effect exists in a real system and can help damp the oscillation of
the current observed in Figs.~\ref{fig:fig2} and \ref{fig:fig3}.
%Our proposed mechanism does not exclude the thermally induced hopping 
%mechanism proposed in Ref.~\cite{lewi} but offers an alternative one.
%The real electron transfer process will be a result of competition among
%the 1D chain superexchange, thermally induced hopping, and the partly 
%2D network superexcange 
%different mechanisms.

In summary, we propose a new mechanism for the charge transfer through a 
DNA duplex chain. It is different from the previously proposed 
thermally-induced hopping mechanism in explaining the observed weak 
distance dependence when the number of (T:A) base pairs between (G:C) base 
pairs is larger than ${\cal M}_c$, in that we treat the system fully 
quantum mechanically and emphasize the importance of inter-strand coupling 
between the two strands of the DNA duplex. We found that the series of (T:A)
base pairs in long (G:C)(T:A)$_{\cal M}$(G:C)$_3$
DNA duplex chains is still a quantum tunneling barrier.
The holes in the left (G:C) base pair tunnel through this two-strand
network superexchange barrier instead of one-strand chain superexchange 
barrier, to the right triple (G:C) base pairs.

We wish to thank Julia Berashevich for many helpful discussions.
The work has been supported by the Canada Research Chair
Program and a Canadian Foundation for Innovation (CFI) Grant.

\end{document}